# Compositionally-modulated Si$_{1-x}$Ge$_x$ multilayers with cross-plane thermal conductivity below the thin-film alloy limit

Peixuan Chen[1,2*], N. A. Katcho[3], J. P. Feser[4], Wu Li[3], M. Glaser[2], O. G. Schmidt[1], David G. Cahill[4], N. Mingo[3*], A. Rastelli[1,2*]

[1]Institute for Integrative Nanosciences, IFW Dresden, Helmholtzstr. 20, 01069 Dresden, Germany.

[2]Institute of Semiconductor and Solid State Physics, Johannes Kepler University Linz, Altenbergerstr. 69, A-4040 Linz, Austria.

[3]LITEN, CEA-Grenoble, 17 rue des Martyrs, 38054 Grenoble CEDEX 9, France.

[4]Department of Materials Science and Engineering, Materials Research Laboratory, University of Illinois at Urbana-Champaign, Urbana, Illinois 61801, USA.

**\*e-mail:** peixuan.chen@jku.at; natalio.mingo@cea.fr; armando.rastelli@jku.at

**Abstract**

We describe epitaxial Ge/Si multilayers with cross-plane thermal conductivities which can be systematically reduced to exceptionally low values, as compared both with bulk and thin-film SiGe alloys of the same average concentration, by simply changing the thicknesses of the constituent layers. *Ab initio* calculations reveal that partial interdiffusion of Ge into the Si spacers, which naturally results from Ge segregation during growth, plays a determinant role, lowering the thermal conductivity below what could be achieved without interdiffusion (perfect superlattice), or with total interdiffusion (alloy limit). This phenomenon is similar to the one previously observed in alloys with embedded nanoparticles, and it stresses the importance of combining alloy and nanosized scatterers simultaneously to minimize thermal conductivity. Our calculations thus suggest that superlattices with sharp interfaces, which are commonly sought but difficult to realize, are worse than compositionally-modulated $Si_{1-x}Ge_x$ multilayers in the search for materials with ultralow thermal conductivities.

The discovery [1-5] that the thermal conductivity (κ) of alloys is much lower than that of their pure components spurred the development of solid-solution-based thermoelectric materials. A renewed stimulus has been the realization that the alloy κ can be further lowered by embedding nanodots [6]. Theoretical studies have identified the simultaneous presence of alloy disorder and nanoscale scattering mechanisms, acting on a wider range of phonon wavelengths, as the reason for the diminished κ [7-9]. More in general, demonstrations of low κ in bulk materials suggest that scatterings at multiple length-scales are at play [10, 11].

From a fundamental point of view, it is important to unequivocally demonstrate and quantify these phenomena on well controlled systems, such as epitaxial superlattices. Specifically, one question is whether the κ of a superlattice can be further reduced by the simultaneous action of alloy scattering. For SiGe-based superlattices, which have been extensively investigated in the past [12-22] and are the focus of this work, even the issue whether their κ can be lower than the corresponding homogeneous alloy is still debated [23].

From an applied point of view, it is also important to elucidate whether less Ge could be used if the homogeneous alloy is replaced by a nanostructured alloy of similar κ, thus reducing materials cost (Ge is 100 times more expensive than Si [24].) Here we show that the answer to the two above questions is "yes". By combining extensive experimental results and *ab initio* calculations, we argue that partial interdiffusion in superlattices, which naturally occurs during their fabrication, results in lower κ values than those which may be achieved in the absence of interdiffusion, and with homogeneous alloys. We also show that superlattices with an average Ge concentraction of just 5% can yield $\kappa < 8$ W/m-K, and that 12% Ge brings $\kappa$ below 3 W/m-K. For comparison, homogeneous SiGe bulk alloys require concentrations above 20% Ge to reach their lowest $\kappa$ value of ~8 W/m-K. Finally, the conclusion that compositionally modulated superlattices can have $\kappa$ values lower than the corresponding alloys [23, 25] remains valid even taking into account the reduction of $\kappa$ produced by phonon boundary scattering in thin films.

The samples used for this work are $(Si)_m/(Ge)_n$ multilayers grown by molecular beam epitaxy on Si(001) wafers at a substrate temperature of 500°C. They consist of $N$ periods of $n$ monolayers (ML) of Ge separated by $m$ ML of Si, as sketched in Fig. 1**a**. The growth rates for Si and Ge were 1.0 and 0.06 Å/s (corresponding to 0.7 and 0.04 ML/s), respectively. All parameters were systematically varied in a wide range to reach a clear picture of their impact on thermal conductivity. The actual multilayer thicknesses of selected samples were quantified by x-ray diffraction. Samples can be considered dislocation-free unless otherwise stated.

The sample surface was characterized by atomic force microscopy (AFM). AFM images for samples with $m=43$ and $n=4$, and $n=4.7$ are shown in Figs. 1**c**-**d**, respectively. Since the critical thickness for the formation of three-dimensional (3D) Ge nanodots on Si(001) is about 4.4 ML [26], a planar multilayer without nanodots is obtained for $n=4$, see Fig. 1**c**. Sparse 3D Ge nanodots (also called "hut clusters" [27]) are instead observed for

$n$=4.7. Their density increases from ~3 to ~60×10$^9$ cm$^{-2}$ as $n$ increases from 4.7 to 5.5. By fixing the Ge amount to 3 ML (well below the critical thickness for dot formation) and reducing the Si spacer thickness, we observe that nanodots form when $m$<10 (shown in Figs. 1**e, f**). This phenomenon is ascribed to Ge surface segregation [28-31]. During Si overgrowth of a Ge layer, Ge is incorporated into the growing crystal only gradually because of its lower surface energy compared to that of Si. This means that (i) a nominal (Ge)$_n$ layer is depleted of Ge, (ii) the overlying (Si)$_m$ layer is enriched in Ge, and that (iii) excess Ge is present on the growth front after the completion of a Si layer. (iii) leads to an apparent reduction of the critical thickness for dot formation, while (i) and (ii) imply that our multilayers consist of compositionally modulated Ge$_x$Si$_{1-x}$ alloys. Figure 1**b** shows the expected concentration profiles for multilayers with increasing Ge thickness $n$ and constant Si thickness $m$, calculated using the segregation model proposed by Godbey and Ancona [30], which is based on experiments performed under conditions similar to ours. The model assumes planar growth and does not include the effect of nanodots.

The cross-plane thermal conductivity κ of the multilayers was measured independently with the differential 3ω method [32, 33] and with time-domain thermoreflectance (TDTR) [34, 35] at room temperature (see Supplementary material for details). Figure 2**a** shows the thermal resistance $R_{tot}$ of multilayers with variable Si-spacer thickness ($m$) and fixed thickness of the Ge "phonon barriers" ($n$=3 ML) and period number ($N$=21). $R_{tot}$ increases slowly when $m$>~10, indicating that the Si spacer has only a minor effect on the thermal response of the structure. This finding is in agreement with previous results obtained on nanodot multilayers with $m>22$ [36] and implies that the multilayer can be seen as a sequence of $N$ thermal resistors each with a resistance $R_{barrier}=R_{tot}/N$. Since $\kappa \sim N \cdot (n+m)/R_{tot}$, the thermal conductivity can be tuned by simply varying the Si spacer thickness, as shown in Fig. 2**b**. It is obvious that $\kappa$ cannot be indefinitely decreased with this approach (in the limit of $m=0$ we expect $\kappa$ to coincide with the value of a Ge film). In fact we see $R_{tot}$ starts decreasing when $m<\sim 10$ and that $\kappa$ tends to saturate to a very low value of about 1.2 W/m K (Fig. 2**b**).

In contrast to the weak dependence of $R_{tot}$ on the Si amount per period (Fig. 2**a**), a much stronger dependence on the thickness of the Ge "barriers" and on the number of periods $N$ is observed in Fig. 2**c**, which shows the values of $R_{tot}$ as a function of $n$ for the $(Ge)_n/(Si)_{43}$ multilayers with various values of $N$. It is important to note that, for all samples studied so far, $R_{barrier}$ (Fig. 2**d**) and $\kappa$ (Fig. 2**f**) *do not* show any significant dependence on the total film thickness (or period number $N$). This is evident when comparing the samples with $(m,n)=(43,1)$ and $N=21$ and 301. Fig. 2**d** demonstrates that $R_{barrier}$ increases linearly with $n$ (up to about 6 ML), as anticipated in Ref. [36]. By linear fitting of the data we obtain $R_{barrier} \sim 5 \cdot 10^{-10} \times n$ (m$^2$K/W).

To give a measure of the efficiency of Ge in lowering the thermal conductivity of our multilayers, we define a "thermal resistivity" $\rho$ as the thermal resistance per unit length of deposited Ge (see Fig. 2**e**). The average value is ~3.5 m K/W, which is only slightly lower than that obtained in Ref. [36]. For $n>\sim 6$, $R_{barrier}$ saturates and $\rho$ shows a slight drop. Inspection of the surface of a sample with $n=6.5$ indicates the presence of large dislocated superdomes (with surface densities of ~$2.6\times 10^8$ µm$^{-2}$). Because of their partial strain-relaxation, such superdomes act as sinks for the deposited Ge [37], so that the effective thickness of the Ge barriers cannot be increased significantly beyond ~6 ML. The fact that the thermal resistance drops after plastic relaxation (rather than increasing because of the presence of crystal defects) is attributed to depletion of Ge from the coherently strained regions and nanodots.

Figure 2**f** shows that the thermal conductivity of the $(Ge)_n/(Si)_{43}$ multilayers decreases from ≈19.4 W/ m·K to ≈2.3 W/m·K as $n$ increases from 0.5 to 5.5. Remarkably, the average Ge concentration for these films is very low and varies from ≈1.1% up to only 12%, indicating that this kind of multilayer is very effective in achieving low thermal conductivities compared to alloys.

To understand these results, we computed the thermal conductivity κ by solving the Boltzmann transport equation (BTE) from first principles. In previous publications, some of us have demonstrated the method as applied to embedded nanoparticles [9, 38]. In the present case, calculation of the scattering rates is complicated by the particular geometry of the scatterers. The scattering term in the Hamiltonian consists of a mean mass density profile, M(z), which only varies in the growth direction *z*, plus a randomly fluctuating term in real space, ΔM(x,y,z), corresponding to the local differences between the mass of each atom and M(z) at that location. In the ideal case where no Ge segregation nor diffusion exists, M(z) has an abrupt profile corresponding to either the Si or Ge mass, and the fluctuating term ΔM is zero everywhere. For that case, one can compute the phonon scattering cross sections and scattering rates due to the barrier, exactly to all orders, using the T-matrix formalism (see supplementary materials.)

When segregation leads to a non abrupt profile, the ΔM needs to be included. The random character of ΔM in the x and y directions justifies the neglect of crossed terms containing M(z)ΔM, due to phase cancellation. Then the scattering probability becomes the sum of the separate contributions from the mean barrier profile, and the average local fluctuation contributions integrated over the period of the multilayer. In the limit of constant M(z) (complete interdiffusion), we retrieve the homogeneous alloy scattering rate [9]. In addition to these harmonic scattering sources, the contribution of anharmonic scattering due to 3-phonon processes is included via the Mathiessen rule. This approach permits us to compute the thermal conductivity of realistic Ge concentration profiles.

Simplistically, one might expect that an abrupt barrier profile led to the largest thermal resistance, due to the higher acoustic mismatch at the interfaces. However, the calculation for this case shows otherwise. Whereas $R_{tot}$ calculated for thin barriers (*n*≤2 ML) is in agreement with experimental values, the one for thicker barriers ($n>2$ ML) is considerably below the experiment, settling to a constant value instead of continuing to increase with increasing $n$ (Fig. 3**a-b**, blue lines). Qualitatively this saturation has to do with the barrier acting as an interferometer: when a sharp homogeneous barrier becomes much thicker than the phonon wavelength, the averaged transmission in one dimension can be written in terms of the individual transmissions of its two interfaces, independent of the thickness [39].

The first-principles results are quite different if, instead of sharp barriers, the estimated Ge distribution profiles from Fig. 1**b**, including segregation, are employed (purple dot-dashed curves in Figs. 3**a-b**). In this case, the resistance keeps increasing, and saturates only after $n=5$ ML. Furthermore, the computed results are now in good agreement with the measured ones.

Thus, a very dilute amount of Ge (less than 5% on average), diffused into the Si spacers, can considerably decrease κ below the pure barrier case, and also below the homogeneous alloy case. The physical reason for this effect is that barriers scatter phonons at low frequencies more efficiently than alloy mass disorder, and vice versa. The simultaneous effect of these two mechanisms results in enhanced scattering at all frequencies, yielding a smaller κ. An analogous phenomenon has been shown to take place in alloys with embedded nanoparticles [6-9, 38].

The fixed 3ML thickness with variable period case has also been calculated from first principles, for the corresponding expected concentration profiles (Fig. 3**c**) Good agreement with the experimental results is obtained for Si spacer lengths of 1.5 nm and above. For periods below 1.5 nm, multiple layer interference effects would need to be included in the calculation. In addition, these shorter period samples begin to show the presence of nanodots, which would need to be treated differently than the planar barriers considered here. Developing such a model is beyond the scope of this paper.

We now compare the $\kappa$ values of our $(Si)_m/(Ge)_n$ multilayers and that of $Si_{1-x}Ge_x$ alloys with the same average Ge concentration ($x \sim n/(n+m)$). Figure 3**b** shows that the $\kappa$ values of multilayers (both with and without interdiffusion) are substantially lower than the $\kappa$ values of the corresponding *bulk* alloys. This is further illustrated by the plot of Fig. 4**a**, which displays the $\kappa$ values of our multilayers together with those of bulk alloys[5, 40] and also previous experimental reports on SiGe superlattices [12-21]. Not only do our superlattices lie below the alloy values, but also below most of the previous data, with the exclusion of those corresponding to structures with high density of crystal defects. As recently stressed in Ref. [23], the comparison between thin-film measurements and bulk values may be misleading, as $\kappa$ may depend not only on $x$ but also on the total film thickness [25]. In fact, it was shown that $\kappa$ for a $Si_{1-x}Ge_x$ alloy *thin-film* with $x \sim 20\%$ and thickness of ~39 nm can be as low as ~1.8 W/m K because of phonon scattering at the interface between film and substrate (see dashed curve and green datapoints in Fig. 4**b**).

To test whether our multilayers can beat also the "*thin-film* alloy limit" we have designed $(Si)_{13}/(Ge)_3$ multilayers ($x$~19%) with $N$=51 and 101. Their thermal conductivities are shown in Fig. 4**b** as a function of film thickness together with all other samples presented in this study. In particular, for the sample with $N$=101 (which is still pseudomorphic), κ is about three standard deviations below the values reported for *dislocated* alloy films in [23], confirming that the reduction of κ does stem from multilayered structure and not from the limited film thickness.

In conclusion, we have achieved multilayered Si/Ge structures with tunable thermal conductivity values well below the alloy limit. The observed dependences of κ on barrier and period thicknesses carry the signatures of partial interdiffusion of Ge from the barriers into the Si spacers driven by segregation. *Ab initio* simulations demonstrate the crucial role of interdiffusion in lowering the conductivity below what could be expected for superlattices with abrupt interfaces. This phenomenon is related to the previously observed reduction in the thermal conductivity of alloys with embedded nanoparticles, and it underlines the important of combining scattering mechanisms acting on different phonon wavelengths to achieve the lowest possible thermal conductivity. This approach may be relevant to many applications requiring low κ, and in particular to the development of on chip superlattice thin film thermoelectrics [41].

We are grateful to V. Fomin, D. Nika and A. Cocemasov for fruitful discussions, J. J. Zhang, D. J. Thurmer and G. Chen for assistance with the MBE growth, T. Etselstorfer and J. Stangl for x-ray diffraction measurements, I. Daruka for fruitful discussions on the calculation of Ge profiles, D. Grimm, I. Mönch, A. Halilovic, G. Katsaros, U. Kainz, S. Bräuer for assistance with processing samples for 3ω measurements, D. A. Broido, D. A. Stewart, and L. Lindsay for fruitful discussions on the *ab initio* approach. We thank P. Hopkins for sharing a preprint of ref. [23]. This work was supported by the SPP1386 (RA 1634/5-1), EU FP7 (NEAT, Grant No. 263440), CEA (THERMA), ANR-08-NANO-P132-48 (ACCATTONE), and the EFRE program. TDTR measurements at the U. Illinois were supported by the US Air Force Office of Scientific Research (AFOSR) Multidisciplinary University Research Initiative Grant No. FA9550-08-1-0407.

# Figure 1

**Figure 1:** (a) Schematic illustration for $(Ge)_m/(Si)_n$ multilayered structure; (b) expected concentration profiles of a single period for different Ge layer thicknesses $n$ and fixed Si layer thickness $m=43$, employing the model of Ref. [30]. (c)-(f) AFM images of topmost layers of Ge/Si multilayers obtained by deposition of the indicated numbers ($m, n$) of Si and Ge monolayers. The number of Ge periods $N$ is 21.

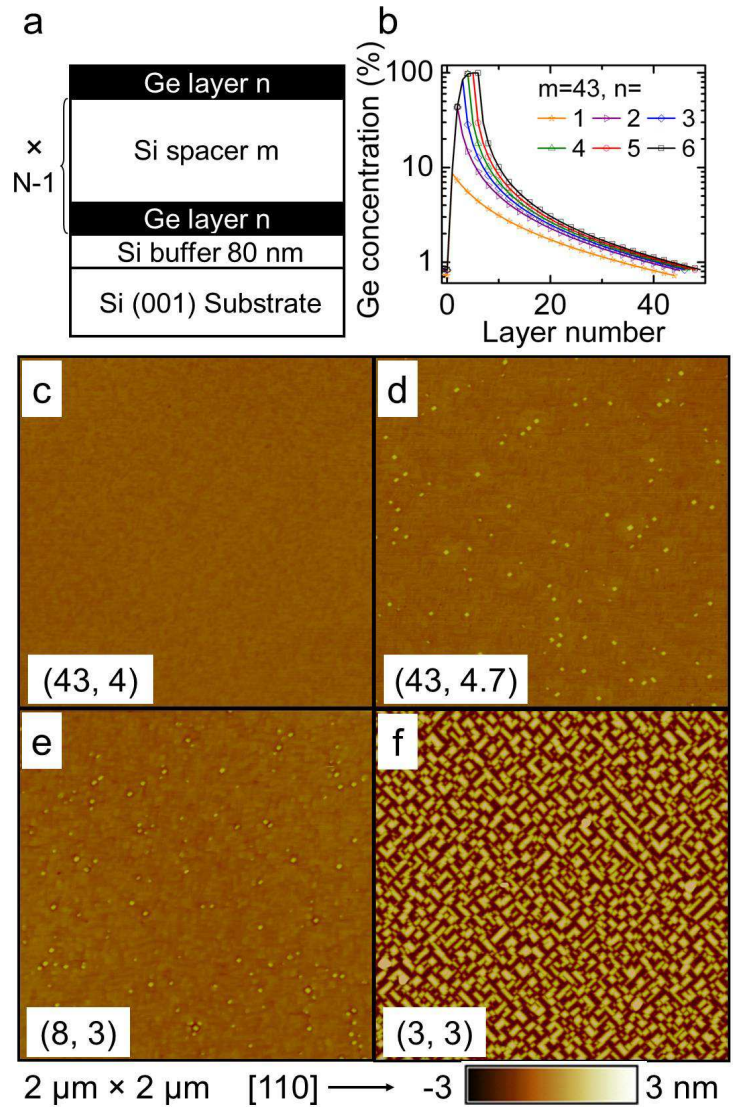

# Figure 2

**Figure 2:** (a) Total thermal resistance $R_{tot}$ of Ge/Si multilayers with fixed Ge amount per layer ($n$=3 ML) and variable Si spacer thickness $m$. The Ge period number $N$ is 21. (b) corresponding thermal conductivity values, (c) total thermal resistance, (d) barrier thermal resistance, (e) thermal resistivity, i.e. thermal resistance per unit length of Ge, and (d) thermal conductivity $\kappa$ as a function of Ge layer thickness of multilayers with varying Ge barrier thickness and fixed Si spacer thickness ($m$=43 ML). The red and blue symbols refer to the results measured by 3ω and TDTR, respectively. The green and yellow dash lines indicate regions of planar multilayers, nanodot-multilayers and multilayers with dislocated nanodots (superdomes).

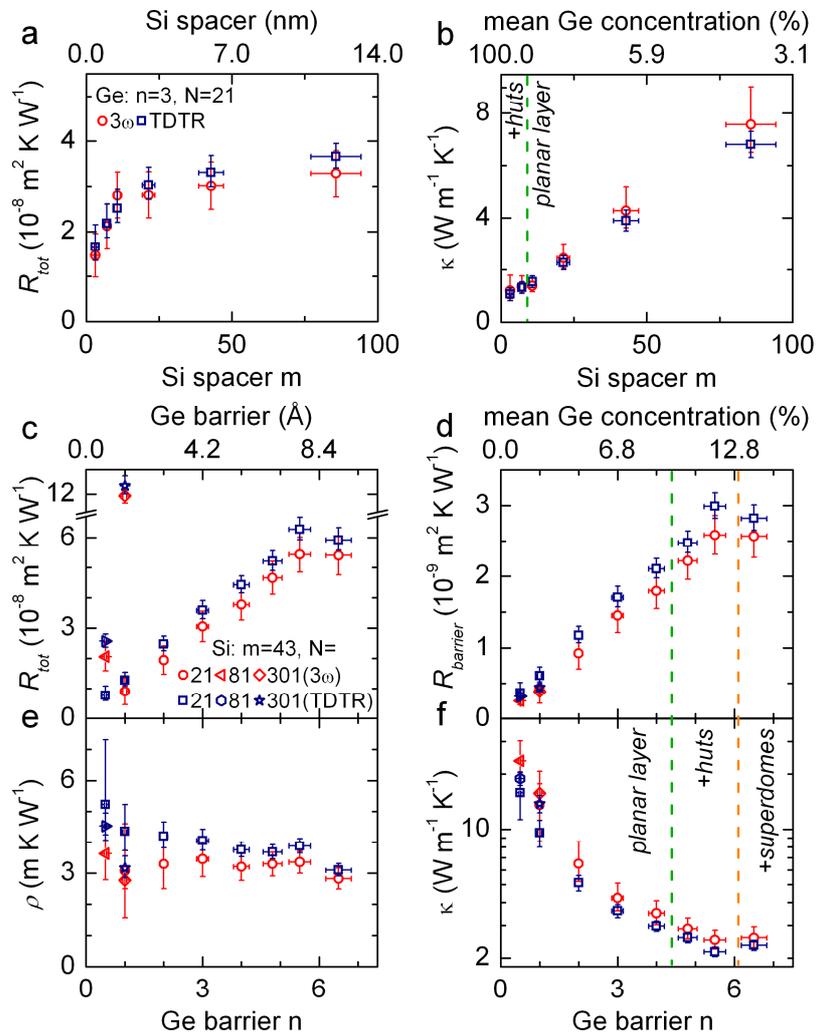

# Figure 3

**Figure 3.** (a,b) Comparison of experimental and theoretical κ and total thermal resistance $R_{tot}$ for different Ge barrier thicknesses $n$, with a constant Si spacer thickness $m=43$. (c) Comparison of experimental and theoretical thermal conductivity as a function of Si spacer thickness $m$, for a constant Ge barrier equivalent thickness $n$ of 3 ML.

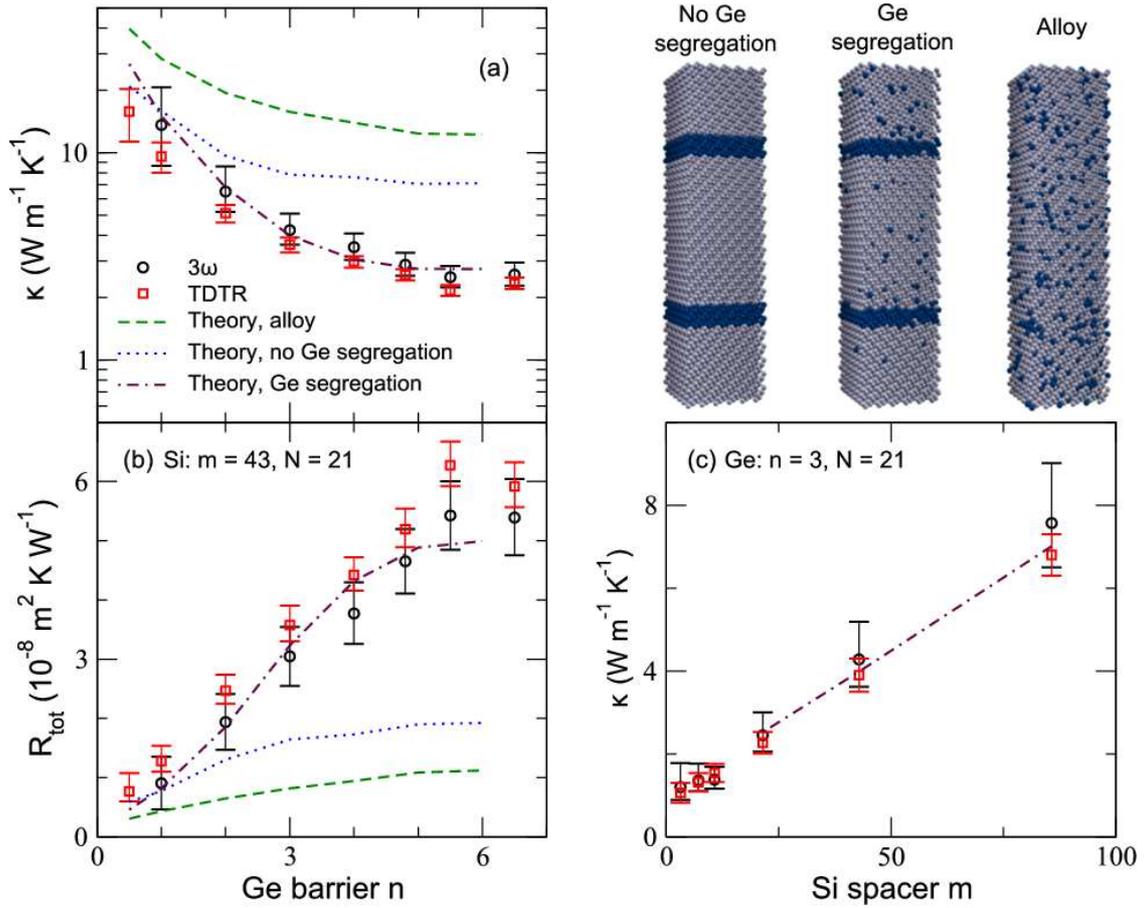

# Figure 4

Figure 4. Thermal conductivity of the grown $(Si)_m/(Ge)_n$ multilayers as a function of (a) mean Ge concentration, (b) total film thickness. The thermal conductivities are obtained by averaging the values of the $3\omega$ and TDTR results. In fig. (b), the mean Ge concentration $x \sim n/(n+m)$ is also indicated near each datapoint. For comparison, bulk alloys and multilayers in ref. [5, 12-21, 40] are shown in (a), the *thin-film* alloy limit measured in ref. [23] is also shown in (b).

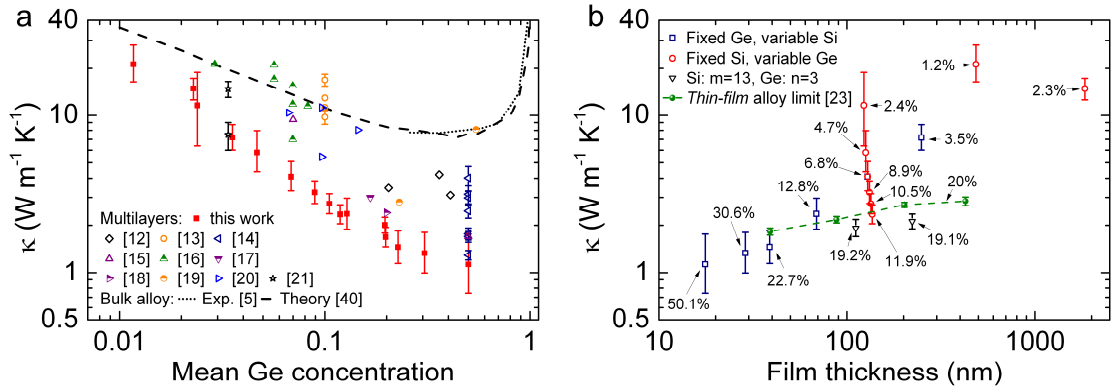